\documentstyle[preprint,eqsecnum,aps]{revtex}
\begin{document}
\tightenlines
\draft

\title{Hairs on the cosmological horizon}

\author{Rong-Gen Cai\footnote{
Electronic address: cairg@ctp.snu.ac.kr}
}
\address{Center for Theoretical Physics,
Seoul National University, Seoul  151-742, Korea}
\author{
Jeong-Young Ji\footnote{
Electronic address: jyji@phyb.snu.ac.kr}
}
\address{Department of Physics Education,
Seoul National University, Seoul  151-742, Korea}

\maketitle

\begin{abstract}

 We investigate the possibility of having hairs on the cosmological
horizon. The cosmological horizon shares similar properties of black
hole horizons in the aspect of  having hairs on the horizons.
For those theories admitting haired black hole solutions, 
the nontrivial matter fields may  reach and extend beyond 
the cosmological horizon.   For Q-stars and boson stars, the matter  
fields cannot reach the cosmological horizon. The no short hair
conjecture keeps valid, despite the asymptotic behavior (de Sitter
or anti-de Sitter) of black hole solutions. We prove  the no scalar 
hair theorem for anti-de Sitter black holes. Using 
 the Bekenstein's identity method,  we also  prove the 
no scalar hair  theorem for the de Sitter space and de Sitter black 
holes if the scalar potential is convex.

\end{abstract}

\pacs{PACS numbers: 04.70.Bw, 04.40.-b}

\section{Introduction}

There  are two no  hair conjectures   in the gravitational  physics.
Although they  have not  been proven  rigorously,  they are  often
referred to as  no hair theorems.  One  is the cosmic  no hair
theorem \cite{Hawking}, which states as  follows \cite{MS}:  Any 
 solution   of the Einstein equations with a positive cosmological  
constant that (i) accepts a  synchronous coordinate  system, 
(ii)  has a   non-positive three-curvature, (iii)  has an  
energy-momentum  tensor satisfying
the   strong  and   dominant   energy  conditions,   will   become
asymptotically de Sitter (at least on patch).  This would imply that
the inflation  of the  universe is  a  natural phenomenon  that can
explain the isotropy and homogeneity seen today in the universe.

The other  is the  no hair  theorem of   black 
holes \cite{Ruffini}. It  is generally believed that the collapse  
of a massive body  will finally lead to the formation of a black hole
and  the external gravitational field of the black hole  settles down 
to the  Kerr-Newman solution of the Einstein-Maxwell equations, 
specified by only three parameters: mass, electric (and/or magnetic)
charge, and  angular momentum \cite{Israel}. This  theorem 
 indeed excludes   scalars \cite{Chase}, massive  
vectors \cite{Bek1}, spinors \cite{Hart}, and Abelian Higgs hairs 
(Maxwell-complex Higgs scalars) \cite{Adler}  from
a stationary black hole exterior. However,  this situation  has been 
changed dramatically since   the    discovery   of   colored    black 
holes    in   the Einstein-Yang-Mills theory \cite{Bizon} 
in 1990. Since then, a lot of black  holes  with  different hairs have 
been   found (see, e.g., Ref.\cite{NQS}).

With the discovery of haired  black holes, naturally much attention
has been drawn to reexamine  the no hair theorem  of black holes.
de Alwis \cite{Alw}  discussed the  validity of the  old no  hair
theorem   in  the   stringy   black  holes.   Using   a   conformal
transformation, Saa \cite{Saa}  showed  the  nonexistence of scalar
hairs in a large  kind of theories. In  the Einstein-conformal scalar
field  theory,  as   is  well   known,  there  exists  the so-called 
Bekenstein black hole solution \cite{Bek3}. But the  conformal scalar   
diverges at  the  horizon and the solution is dynamically unstable 
\cite{Bron}. So  Zannias \cite{Zan}  proved  that  
black hole  horizon  cannot support the  conformal scalar hair  
in the sense  of no hair
theorem. Furthermore, Sudarsky and Zannias \cite{SudZan} recently 
showed that the stress-energy tensor is ill-defined and the
Einstein equations do not hold at the horizon. And hence  
the Bekenstein solution fails to represent a genuine black
hole solution. In Ref.\cite{Bek} Bekenstein proposed a  novel
``no-scalar-hair theorem''   of black   holes, which   rules out   a
multicomponent scalar  field  dressing of   any asymptotically  flat,
static, spherically symmetric black  holes. This theorem  also holds
for scalar-tensor  gravity. Further  Mayo and   Bekenstein~\cite{Mayo}
 investigated   this theorem  for the   charged self-interacting
scalar field  coupled to  an Abelian  gauge field,  or non-minimally
coupled to gravity. In addition, Sudarsky~\cite{Sud}  suggested a
very simple proof of the no hair theorem in the Einstein-Higgs  theory.
N\'u\~nez, Quevedo, and Sudarsky (NQS)~\cite{NQS} made important
progress in understanding the haired black holes by showing that
black holes   have no   short hair:  Under  some  conditions  they
assumed, the   region with  nontrivial  structure of   the nonlinear
matter fields must extend  beyond 3/2 the  horizon radius of black
holes. More issues related  to the no hair  theorem and uniqueness
theorem of black holes can be found in Ref. \cite{Heu}.

When a positive (negative) cosmological constant is present, 
it is widely believed that the Kerr-Newman solution to the 
Einstein-Maxwell equations will become the Kerr-Newman-(anti-)de 
Sitter solution, whose spacetime is the asymptotically (anti-)de 
Sitter one. But the uniqueness theorem for this solution is still 
lacking, contrary to the Kerr-Newman solution.
Further the presence of a positive cosmological constant will
usually accompany the occurrence of a cosmological horizon. Thus
the cosmological constant changes  greatly the asymptotic behavior
and structure of spacetimes.   However all proofs of the no hair 
theorem have been carried out on  the  assumption that   the black hole  
 spacetime is  asymptotically flat. It  is therefore of some 
 interest to  investigate
the effects of the cosmological constant on the no hair theorem and
no short hair conjecture of NQS. The cosmological horizon has
 many similar properties of black hole horizon. For example, 
like the black hole  horizon, the cosmological horizon has
 the Hawking evaporation and the  entropy associated with the
 horizon \cite{Gibbons}, and is  classically stable \cite{Cai}. 
Has  the cosmological horizon the similar property 
in having hairs on the black hole horizon?
On the other hand, the regular solutions such as Q-stars, boson stars,
and gravitational solitons may  be surrounded by a cosmological 
horizon when a positive cosmological constant is present. 
Can these nontrivial matter fields reach and extend beyond 
the cosmological horizon?  In the present work we try to make some 
investigations.

The organization of  this paper is  as follows. In  Sec. II we
consider the Einstein-Yang-Mills theory with a cosmological 
constant. For the regular Bartnik-McKinnon \cite{BM} soliton 
solution surrounded by a cosmological horizon, we find that the 
nontrivial Yang-Mills field may reach and extend beyond 
the cosmological horizon. As the case in the asymptotically 
flat spacetime, to have the required asymptotic behavior the
 nontrivial  Yang-Mills field also must extend beyond a critical point
 satisfying $r=3m(r)$, where $m(r)$ is a mass function in the metric.
In Sec. III we extend to discuss the hairs on the
cosmological horizon in those theories allowing haired black holes,
and investigate the effect of cosmological constant on the no short 
hair conjecture of black holes. For the Q-stars and boson stars,
however, we find that the matter fields cannot reach the cosmological 
horizon.   In Sec. IV we discuss the no scalar hair theorem for 
asymptotically (anti-)de Sitter black holes. The 
conclusion and discussion are given in Sec. V.

\section{Einstein-Yang-Mills theory with a cosmological constant}

It is the Einstein-Yang-Mills theory in which Bartnik and 
McKinnon \cite{BM} first found the nontrivial 
gravitational soliton solution and subsequently some authors 
\cite{Bizon} discovered 
numerically the first haired black hole. Here the meaning of  ``hair'' 
follows Ref. \cite{NQS}: In a given theory, there is black hole hair
when the spacetime metric and the configuration of the other fields 
of a stationary black hole solution are not completely specified by 
the conserved charges defined at asymptotic infinity. 
In  this section we discuss
the Einstein-Yang-Mills theory with a cosmological constant. 
For a positive cosmological constant, more recently, many authors 
\cite{Torri,Volkov} have
investigated the system. Due to the nontrivial
asymptotic behavior some new phenomena have been revealed.

The action of the Einstein-Yang-Mills theory with a cosmological
constant is
\begin{equation}
S=\frac{1}{16\pi} \int d^4x\sqrt{-g}[R-2\Lambda 
     -\frac{1}{g^2}{\rm Tr} F^2],
\label{e1}
\end{equation}
where $R$ is the scalar curvature, $\Lambda$ is the cosmological
constant,  $F$ is the SU(2) Yang-Mills field strength and $g$ is the
coupling constant of the field. Throughout this paper the units $G=c=1$
have been used. We now consider the solution whose metric is of 
the form 
\begin{equation}
ds^2=-\mu (r)e^{-2\delta(r)}dt^2 +\mu^{-1}(r)dr^2 +r^2d\Omega ^2,
\label{e2}
\end{equation}
where
\begin{equation}
\mu(r)=1-\frac{2 m(r)}{r}-\frac{1}{3}\Lambda r^2,
\label{e3}
\end{equation}
$m(r)$ denotes the mass function and $d\Omega^2$ represents 
the line element on the unit 2-sphere. Throughout this paper,
we require that the solution  is asymptotically  de Sitter
when $\Lambda >0$ \footnote{The authors in \cite{Volkov} found that in
the Einstein-Yang-Mills theory with a positive cosmological constant
there are not only the asymptotically de Sitter solution, but also the
so-called the bag of gold solution and the compact regular solution with 
space topology $S^3$, depending on the cosmological constant and the
node number of the Yang-Mills amplitude. It would be a quite interesting 
subject to further investigate the  latter two asymptotic behaviors from 
the point of view of the no hair conjecture. Here we restrict ourselves 
to the case of the first asymptotic
behavior.} or anti-de Sitter for $\Lambda <0$ \footnote{In fact, when
$\Lambda <0$, the so-called topological black holes, whose topology of 
event horizons is no longer the  2-sphere $S^2$, may appear. 
 In this paper the topology of event horizons is restricted to
the 2-sphere.}. For the 
metric~(\ref{e2}) we have 
\begin{equation}
 \lim _{r \rightarrow \infty} m(r)=M, \  \
 {\rm and}\ \ \lim _{r \rightarrow \infty} \delta (r)=0,
\label{e4}
 \end{equation}
where $M$ is a constant. In the solution (\ref{e2}), when $\delta (r)$ 
and $m(r)$ vanish, the line element (\ref{e2}) describes the de Sitter
space ($\Lambda >0$) or anti-de Sitter space ($\Lambda <0$). 
For the de Sitter space the future infinity is spacelike. 
This means that for each observer moving 
on a timelike world line there is an event horizon separating the region 
of spacetime which the observer can
never see from the region that he can see
if he waits long enough. In other words, the event horizon is the boundary
of the past of the observer's world line. This event horizon is called a 
cosmological event horizon. It is located at the coordinate singularity 
$ r_c=\sqrt{3/\Lambda}$ in the solution (\ref{e2}). When $\delta (r)=0$,
$m(r)=M$ is a constant and $\Lambda>0$ in (\ref{e2}), the solution is just
the Schwarzschild-de Sitter spacetime. If $9\Lambda M^2 <1$, the equation
$\mu(r)=0$ then has two positive roots. The large one is just the 
cosmological
horizon, beyond which $\mu(r) <0$, while the small one is the black hole
horizon. When $\Lambda <0$, the cosmological horizon is absent. For
more details about the cosmological and black hole horizons, the reader 
is referred to Ref.\cite{Gibbons}. For the solution we are considering
in the action (\ref{e1}), the black hole horizons and /or a cosmological
horizon of the metric (\ref{e2}), if exist, 
are regular and hence the metric functions $m(r)$, $\delta (r)$ 
 are finite on the horizons. Further we require that the matter fields
 are also finite on the horizons.

For the Yang-Mills gauge potential, we take the following ansatz
\begin{equation}
A=w(r)\tau _1 d\theta +(w(r)\tau _2 +\cot \theta \tau _3)\sin \theta
d\phi,
\label{e5}
\end{equation}
where $\tau _i$ ($i=1,2,3$) are three  Pauli matrices. 
In the metric (\ref{e2}),
we have equations of motion
\begin{eqnarray}
&& m'(r)=\mu \frac{w'^2}{g^2}+\frac{2V(w)}{g^2r^2}, 
\label{e6} \\
&& \delta'(r)=-\frac{2w'^2}{g^2r},
\label{e7} \\
&& r^2e^{\delta}(\mu e^{-\delta}w')'=
\frac{\partial V(w)}{\partial w},
\label{e8}
\end{eqnarray}
where
\begin{equation}
V(w)=\frac{1}{4}(1-w^2)^2,
\label{e9}
\end{equation}
and  a prime denotes derivative with respect to $r$.  In 
Eqs.~(\ref{e6})-(\ref{e8}) there exist two trivial 
exact solutions with horizons.  One is the Schwarzschild-(anti-)de 
Sitter solution when the Yang-Mills 
potential $w=\pm 1$. The solution includes of course the (anti-)de
Sitter space  as a special case, that is, $\delta (r)=0$ and $m(r) =0$ 
in the Eq. (\ref{e2}). When $w=0$, the exact solution is the 
Reissner-Nordstr\"om-(anti)-de Sitter spacetime.  
The authors in \cite{Volkov} also gave  a few other exact solutions for
a positive cosmological constant. Now we want to
discuss the nontrivial solution with horizons. Here the word 
``nontrivial'' means that the  Yang-Mills 
potential $w$ is no longer a trivial constant throughout the whole  
spacetime.

 From Eq.~(\ref{e8}) and by using Eqs.~(\ref{e6}) and (\ref{e7}), 
we can obtain
\begin{equation}
r^2 \mu w''+2w'\left ( m-\frac{1}{3}\Lambda r^3
-\frac{2V(w)}{g^2r}\right) =\frac{\partial V(w)}{\partial w}.
\label{e10}
\end{equation}
Multiplying Eq.~(\ref{e10}) by $w'$, one has
\begin{eqnarray}
&& \left (\frac{1}{2}\mu r^2 w'^2\right )'
+2w'^2 \left [-\frac{1}{4}(\mu r^2)'
+m \right. \nonumber \\
&&~~~~~~~~~\left. -\frac{1}{3}\Lambda r^3-\frac{2V(w)}{g^2r}\right ]
 =\frac{\partial V(w)}{\partial w}w'.
\label{e11}
\end{eqnarray}
 Defining 
\begin{equation}
E(r)=\frac{1}{2} \mu r^2 w'^2 -V(w),
\label{e12}
\end{equation}
 it is easy to show
\begin{equation}
E'(r)=-\left [\frac{2}{g^2 r}E +( 3m -r)\right ]w'^2.
\label{e13}
\end{equation}
Furthermore  we can get the following equation
\begin{equation}
\frac{d}{dr}\left(Ee^{-\delta}\right)=-[3 m(r)-r] e^{-\delta}w'^2.
\label{e14}
\end{equation}
We note that this equation is completely identical 
with  Eq. (25) in Ref.~\cite{Sud} 
although our case is that a cosmological constant is present. 
Now we discuss two  cases, respectively.

(i) Bartnik-McKinnon (BM) soliton  surrounded
by a cosmological horizon, say $r_c$.  In this case, the cosmological 
constant is positive and a cosmological horizon appears. The origin 
is regular, that is, the  metric function $m(r)=\delta(r)=0$ and the 
Yang-Mills potential $w=\pm 1$ at the origin. 
Can  the nontrivial Yang-Mills field reach and extend beyond the 
cosmological horizon
under certain conditions? If the nontrivial Yang-Mills field reaches 
the cosmological horizon, we then have $\mu(r_c)=0$, 
$E(r_c)= -V[w(r_c)]<0$ and the function $Ee^{-\delta}$ is negative 
semidefinite at the cosmological horizon. Obviously, 
when $r<3m(r)$, the right hand of side of Eq.~(\ref{e14}) is always 
negative. Thus, to have the asymptotically de Sitter behavior
the nontrivial Yang-Mills field 
must extend beyond the critical point $r_{\rm crit}$ satisfying
\begin{equation}
r_{\rm crit}=3m(r_{\rm crit}).
\label{e15}
\end{equation}
Otherwise, the equation (\ref{e14}) can not be satisfied by a 
nontrivial Yang-Mills field (that is, $w$ is not a trivial 
constant).
Due to the fact that the solution we are considering has only the
cosmological horizon and the black hole horizon is absent, the critical 
point (\ref{e15}) in fact is inside the cosmological horizon. 
Indeed, the data in Ref. \cite{Volkov} showed this fact. 
In other words, there is no obstacle for the nontrivial Yang-Mills field
reaching and extending beyond the cosmological horizon. This was 
already showed numerically in \cite{Volkov}.

(ii) Colored black hole. 
In this case, when the cosmological constant is positive, there 
may exist  not only the black hole horizon, but also the cosmological 
horizon, while only the black hole horizon is present when 
the cosmological constant is negative.  Inspecting
Eq.~(\ref{e14}), we can see clearly that the cosmological constant 
does not appear explicitly. Note that the function $Ee^{-\delta}$ 
is still negative semidefinite 
at the black hole horizon (and cosmological horizon if  $\Lambda >0$). 
The condition (\ref{e15}) remains unchanged in order
to have the correct asymptotic
behavior of solutions. But it should be pointed out that now the critical
point (\ref{e15}) is outside the black hole horizon. 
 That is, if the nontrivial Yang-Mills field reaches  the black hole horizon,
then it must extend beyond the critical point (\ref{e15}). The size of
the hairosphere will be discussed in the next section.
Therefore, for asymptotically (anti-)de Sitter colored black holes, 
the no short hair conjecture of NQS keeps valid.

So far  we have seen that for the asymptotically (anti-)de Sitter 
colored black holes,  the nontrivial Yang-Mills field at the 
black hole horizon must extend beyond a critical point satisfying 
Eq.~(\ref{e15}). In fact, Eq.~(\ref{e15}) is still a universal 
condition for those theories allowing the haired black holes when a 
cosmological constant is introduced. For  Q stars and boson 
stars, however, the matter fields cannot reach the cosmological 
horizon. In the next section, we will discuss the general case.

\section{No short hair conjecture and cosmological constant}

On the basis of the  investigation of 
all black holes with hairs discovered in the different theories
in recent years,  N\'u\~nez,  Quevedo  and  Sudarsky~\cite{NQS}
 have found that the region with nontrivial  structure of
the nonlinear matter fields must extend beyond 3/2 the horizon
radius, being independent of all other parameters in this theory. 
Further they have argued that this is a  universal lower bound for 
asymptotically flat black holes and the matter satisfying the 
following conditions: (i) The weak energy
condition holds; (ii) The energy density $\rho$ falls to zero faster
than $r^{-4}$; (iii) The trace of the stress-energy tensor is
negative. Based on this observation they have put forward the no short
hair conjecture to replace the original no hair theorem.

In this section we would like to show that the presence of the 
cosmological constant does not essentially affects the no short hair
conjecture. We will still work in the spherically  symmetric  
metric~(\ref{e2}). The Einstein equations with a cosmological 
constant are 
\begin{equation}
R_{\mu\nu}-\frac{1}{2}R g_{\mu\nu}  
+\Lambda   g_{\mu\nu}  =8\pi 
T_{\mu\nu},
\label{e16}
\end{equation}
where $T_{\mu\nu}$  represents the stress-energy tensor  of
the matter fields. In the metric (\ref{e2}), 
Eqs.~(\ref{e16}) give
\begin{eqnarray}
&& \delta '(r)=\frac{4\pi r}{\mu}\left(T^t_{~t}-T^r_{~r}\right), 
\label{e17} \\
&& \mu'(r)=r (8\pi T^t_{~t}-\Lambda )+\frac{1-\mu}{r}.
\label{e18}
\end{eqnarray}
The  equation (\ref{e18}) can be rewritten as
\begin{equation}
m'(r)=-4\pi r^2 T^t_{~t}.
\label{e19}
\end{equation}
With the help of the conservation equations of matter fields
$T^{\mu}_{~\nu;\mu}=0$ and Eqs.~(\ref{e17}) and (\ref{e18}), we 
 obtain
\begin{eqnarray}
e^{\delta}(e^{-\delta}r^4T^r_{~r})' &=&\frac{r^3}{2\mu}\left [
(3\mu +\Lambda r^2 -1)(T^r_{~r}-T^t_{~t})  +2\mu T\right ]
\nonumber  \\
&=&\frac{r^3}{\mu}\left [\left(1-\frac{3m}{r}\right) 
\left(T^r_{~r}-T^t_{~t}\right )+\mu T\right],
\label{e20}
\end{eqnarray}
where $T$ denotes the trace of the stress-energy  tensor.
 Now we discuss the Eq.~(\ref{e20}) under the same assumption as 
 in Ref.~\cite{NQS}, 
that is, matter fields satisfy the weak energy condition,
the energy density,  $\rho \equiv -T^t_{~t}$,
goes to zero faster than $r^{-4}$ and the trace of the stress-energy
tensor, $T$, is always negative.
Here the weak energy condition implies that  
$\rho$ is positive  semidefinite and $|T^r_{~r}| \le -T^t_{~t}$. 
 From the regularity of horizons and the finiteness of matter fields 
on the horizons, inspecting (\ref{e20}) we must have
\begin{equation}
T^r_{~r}-T^t_{~t} =0 \ \ {\rm at}\ \ r=R_h,
\label{e21}
\end{equation}
where $R_h$ stands for a cosmological or black hole horizon location.

First we consider the case where  the only cosmological horizon is 
present at $r_c$, that is,  a soliton solution surrounded by a 
cosmological horizon.  For all the theories allowing haired black holes 
such as the Einstein-Yang-Mills theory, Einstein-Skyrme theory, 
Einstein-Yang-Mills-dilaton theory with or without 
an additional  potential term, 
Einstein-Yang-Mills-Higgs theory, Einstein-non-Abelian-Procca theory 
(the quantity $T^r_{~r}-T^t_{~t}$ of 
these theories is given in Ref. \cite{NQS}), 
self-gravitating global monopoles~\cite{Kas} and gauge monopoles~\cite{Lee}, 
one has  
\begin{equation}
T^r_{~r}-T^t_{~t} = \mu P,
\label{ten}
\end{equation}
where $P$ is a positive semidefinite function of $r$.
Hence all these theories satisfy the condition (\ref{e21}) and
then Eq.~(\ref{e20}) becomes
\begin{equation}
e^{\delta}(e^{-\delta}r^4T^r_{~r})' 
= r^3 \left[ \left(1 - \frac{3m}{r} \right) P + T \right] .
\label{easy}
\end{equation} 
Note that the matter fields satisfy the weak energy condition and
Eq.~(\ref{e21}). The function $e^{-\delta} r^4T^r_{~r}$ is
negative semidefinite at the cosmological horizon. 
On the other hand, note that 
 the right hand side of Eq.~(\ref{easy}) is negative
semidefinite if $r<r_{\rm crit}$, where  
\begin{equation}
r_{\rm crit}=3m(r_{\rm crit}), 
\label{e22}
\end{equation}
and the function $e^{-\delta}r^4T^r_{\ r}$ must  asymptotically  
approaches zero as $r\rightarrow \infty$ so that the solution is 
asymptotically de Sitter. Therefore if the nontrivial matter fields
reach the cosmological horizon,  they  must satisfy  the
critical point relation (\ref{e22}).  Thus we  obtain the condition 
(\ref{e15}) for the general case.  
Because the nontrivial Yang-Mills field 
can reach and extend beyond 
the cosmological horizon, we have no
reasons to doubt that other  nonlinear matters mentioned above 
cannot reach the cosmological horizon. Thus we conclude that
the cosmological horizon can support the nontrivial 
nonlinear matter fields.

The equations (\ref{e20}) and (\ref{ten}) and the property of the
cosmological horizon lead to the following theorem.

{\it Theorem 1}: In the spherically symmetric, asymptotically 
(anti-)de Sitter black hole spacetime with matter fields satisfying
the weak energy condition, the energy density going to zero faster than
$r^{-4}$, and the trace of stress-energy tensor being nonpositive, 
if the nontrivial matter configuration reaches  the  black hole horizon,
it must extend beyond a universal critical point satisfying 
$ r_{\rm crit}=3m(r_{\rm crit})$, where $m(r)$ is the mass function 
in the metric. Or the no short hair conjecture keeps valid for
asymptotically (anti-)de Sitter black holes.

{\it Proof}: 
For the asymptotically anti-de Sitter black hole solution,
we have 
\begin{equation}
\mu (r_b)=0, \ \ {\rm and} \ \
\mu(r) >0 \ \ {\rm for}\ \ r>r_b,
\label{e23}
\end{equation}
and $\delta (r)$ is finite at the horizon and approaches zero 
as $r \rightarrow \infty$, where $r_b$ denotes the black hole horizon.
For the asymptotically de Sitter solution, 
we  have
\begin{equation}
\mu (r_b)=\mu (r_c)=0,
\label{e24}
\end{equation}
and 
\begin{equation}
\mu (r) >0 \ \ {\rm for}\ \ r_b<r<r_c,\    \
{\rm and} \ \ \mu(r) <0 \ \ {\rm for}\ \ r>r_c,
\label{e25}
\end{equation}
where $r_c$ represents the cosmological horizon. Note that the function
$e^{-\delta}r^4 T^r_{~r}$ is always negative semidefinite at the black 
hole horizon (and cosmological horizon 
if $\Lambda >0$) and the cosmological constant does not appear
explicitly in Eq.~(\ref{easy}). To have the required asymptotic behavior, 
the function $e^{-\delta}r^4T^r_{\ r}$ must go to zero as $r\rightarrow
\infty$. Therefore the nontrivial  matter fields must extend beyond the
critical point (\ref{e22}), if they reach the black hole horizon.

Here it is worth stressing that from mathematical expressions, our
result (\ref{e22}) is completely the same as the one of NQS. Note that
$m'(r) \ge 0$ because of the positive semidefinite energy density
 and requiring the   asymptotic flatness of  spacetime, they could
 further   write down:
 \begin{equation}
 r>r_{\rm crit}=3 m(r_{\rm crit}) >3m (r_b)=\frac{3}{2}r_b.
\label{e26}
 \end{equation}
Therefore they  can assert  that the nontrivial matter fields must
extend beyond 3/2 the horizon radius of black holes. For our case,
because the horizons are determined by the equation $\mu(r)=
1-2m/r-\Lambda r^2/3 =0$, we cannot obtain the
last equality in Eq.~(\ref{e26}). But in general, we can say that 
 the positive cosmological constant widens the hairosphere, while  
the hairosphere thins for a negative cosmological constant, compared to
the one  in the asymptotically flat black hole solutions
\cite{NQS}.  From our  result, furthermore, we can 
see that the essence of the no short hair conjecture of black holes 
keeps valid for the asymptotically (anti-)de  Sitter black  holes.
In addition,  we would like to point out that 
although the cosmological constant does not come 
explicitly into the critical point relation (\ref{e22}),
the cosmological constant has effects on the black hole hairs.
Adding a positive (negative) cosmological constant corresponds to 
adding a repulsive (attractive) force. If the cosmological constant 
is too large, one would have no equilibrium 
configurations \cite{Volkov}.

In recent years, there has  been considerable interest in
two kinds of stars: Q-stars and boson stars 
\cite{Rev1,Rev2,Rev3}. When a positive cosmological 
constant is added to these theories, it  is  expected that
these regular stars would be surrounded by a cosmological 
horizon. Can the matter fields constructing the star reach 
the cosmological horizon?  Note that a conserved particle number
associated with the Noether current appears always in these theories.
A system of self-gravitating real massive scalar field does not 
admit regular static solutions, since there is no conserved current 
leading to particle number conservation. Consider a general 
complex scalar $\Phi $ with a self-interacting  potential 
$U(\Phi ^*\Phi)$. Its Lagrangian is
\begin{equation}
L_{\rm matter}=-g^{\mu\nu}\partial _{\mu}\Phi ^*
\partial _{\nu}\Phi -U(\Phi^*\Phi),
\label{e27}
\end{equation}
which has a conserved current 
\begin{equation}
j^{\mu}=-i \sqrt{-g}g^{\mu\nu}(\Phi ^*\partial _{\nu} \Phi
        -\Phi \partial _{\nu}\Phi ^*),
\label{e28}
\end{equation}
and a conserved particle number 
\begin{equation}
N=\int d^3 x j^0.
\label{e29}
\end{equation}
The Noether charge prevents the star from diffusion. 
 From (\ref{e27}) the stress-energy  tensor is
\begin{eqnarray}
T_{\mu\nu}&=&\partial _{\mu}\Phi ^*\partial _{\nu}\Phi
        + \partial _{\mu}\Phi \partial _{\nu}\Phi ^*
      \nonumber \\
		  &-& g_{\mu\nu}[ g^{\alpha \beta}\partial _{\alpha}\Phi ^* 
          \partial _{\beta}\Phi +U(\Phi ^*\Phi)].
\label{e30}
\end{eqnarray}
For static, spherically symmetric  configurations, 
the complex scalar field has the form 
$\Phi=\phi (r)e^{-i\omega t}$, where $\omega $ is a nonzero constant. 
In the metric (\ref{e2}), from (\ref{e30}) we have
\begin{equation}
T^r_{~r}-T^t_{~t}=2\mu (r) \phi' (r)^2 +
        2\omega^2 \mu (r) ^{-1}e^{2\delta}\phi (r)^2,
\label{e31}
\end{equation}
from which we see clearly that $T^r_{~r}-T^t_{~t} \ne 0 $ 
unless $\phi (r)=0$  at the cosmological horizon. 
Considering the requirement (\ref{e21}) we conclude that the scalar 
field in the boson stars cannot reach the 
cosmological horizon. Boson soliton stars \cite{Rev2} and 
Q-stars \cite{Col} are two kinds of non-topological soliton
stars. The difference between soliton stars and general boson
stars is that in the absence of gravitational field the 
soliton stars reduce to non-topological solitons. For the general 
boson stars  the theory has  no soliton solution. That is,
the choice $U(\Phi ^*\Phi)$ is very different in the different 
kinds of stars.  But there
are generally the complex scalar fields like $\Phi $ in the 
boson soliton stars and Q-stars in order to have a conserved charge.
These matter configurations in Q-starts and boson soliton stars 
therefore cannot reach the cosmological 
horizon, either, because the potential $U(\Phi ^*\Phi)$ does not
appear in Eq.~(\ref{e31}). In addition, adding  other matter fields
such as, fermion field, Maxwell field, etc. to these theories 
cannot change this result. This point can be seen from Eq.~(\ref{e31}).
In Ref.~\cite{Kas} Kastor and Traschen have argued the possibility 
of having black hole horizons inside various classical field 
configurations by using the condition (\ref{e21}) resulting from 
the Oppenheimer-Volkoff equation of hydrostatic equilibrium.
The Q-stars and boson stars do not allow for black hole horizon 
inside them.  Therefore it further shows that the cosmological horizon 
and black hole horizon share similar  properties from the viewpoint of 
having hairs on them.

\section {No scalar hair theorem and cosmological constant}

Nowadays there are some methods to show that 
for the Einstein-minimally coupled  real scalar field system 
with a positive semidefinite potential, the static, spherically 
symmetric, and asymptotically flat solution is only the 
Schwarzschild solution with a constant scalar field corresponding 
to a zero point of the potential. This just is the no scalar hair 
theorem of black holes \cite{Bek,Sud,Heu}.  In these proofs the 
condition of asymptotic  flatness plays an important role. When 
a cosmological constant is present, this condition is lost. Can one
still prove the no scalar hair theorem? 
In this section we discuss  this problem.

Consider the following action 
\begin{equation}
S=\int d^4x \sqrt{-g}\left 
[ \frac{1}{16 \pi} ( R-2\Lambda) -\frac{1}{2}g^{\mu\nu}
\partial _{\mu}
 \phi \partial_{\nu}\phi -V(\phi) \right ],
\label{e32}
\end{equation}
where  $V(\phi)$  is  a positive   semidefinite 
potential of the scalar field  $\phi$. Varying this action, we  
have the equations of motion 
\begin{eqnarray}
&& R_{\mu\nu}-\frac{1}{2}g_{\mu\nu}R + \Lambda g_{\mu\nu}
\nonumber \\
&& ~~~~~~=8 \pi 
\left[ \partial_{\mu}\phi \partial _{\nu}\phi 
- g_{\mu\nu} \left(\frac{1}{2}(\nabla \phi)^2 +V(\phi)
         \right) \right],
\label{e33} \\
&& \nabla ^2 \phi =\frac{\partial V(\phi)}{\partial \phi}.
\label{e34}
\end{eqnarray}
In  the metric (\ref{e2}) the Einstein equations (\ref{e33})
 reduce to
\begin{eqnarray}
\delta'(r) &=& - 4 \pi r \phi '^2, 
\label{e35} \\
m'(r)   &=& 4 \pi r^2 \left [\frac{1}{2}\mu (r) \phi'^2 
+ V(\phi) \right ].
\label{e36}
\end{eqnarray}
The asymptotic condition  requires that 
\begin{equation}
\lim_{r \rightarrow \infty} \phi'^2 \sim O(1/r^{6+\varepsilon}),
\  \ {\rm and} \ \
\lim_{r\rightarrow \infty} V(\phi) \sim O(1/r^{4+\varepsilon}),
\label{e37}
\end{equation}
where $\varepsilon$ is a positive small quantity. Comparing  with the
case of asymptotic flatness \cite{Sud}, we find that
$\phi'^2$ is required to fall off faster than the one in 
asymptotically flat spacetime.     
The equation of  motion for the  scalar field (\ref{e34})  becomes  
\begin{equation}
(\mu \phi ')' + \left (\frac{2}{r}-\delta '\right ) \mu \phi '
=\frac{\partial V}{\partial \phi}.
\label{e38}
\end{equation}
Following   Sudarsky~\cite{Sud},  we   multiply  Eq.~(\ref{e38}) by 
$\phi'$ to obtain
\begin{equation}
\left [\frac{1}{2} \mu \phi'^2 \right ]' +\left [\frac{1}{2}\mu '
+\left (\frac{2}{r}-\delta'\right) \mu \right ]\phi'^2
=\frac{\partial V(\phi)}{\partial \phi}\phi'.
\label{e39}
\end{equation}
 Let us study the behavior of  solutions for the cases of the 
asymptotically anti-de  Sitter $(\Lambda < 0)$ and  
 the asymptotically 
de Sitter $(\Lambda > 0)$, respectively.

(i) $\Lambda   <0$. In this case, the  solution is required to  be  
asymptotically anti-de Sitter. We have the following theorem.

{\it Theorem 2}: In the   Einstein-minimally coupled   scalar field  
system with  a positive semidefinite scalar potential and   a negative 
cosmological constant, the  static,  spherically symmetric  black hole 
solution with   a regular   horizon and   possessing asymptotically 
anti-de   Sitter  behavior   is  the   Schwarzschild-anti-de  Sitter 
spacetime and the scalar field is a  constant corresponding 
to a local extremum  of  this potential.

{\it Proof}: To prove this theorem, following \cite{Sud} we define 
\begin{equation}
E=\frac{1}{2}\mu \phi'^2 -V(\phi).
\label{e40}
\end{equation}
 it is easy to show 
\begin{equation}
E'(r)=-b\phi'^2,
\label{e41}
\end{equation}
where 
\begin{equation}
b = 4 \pi r E 
- \Lambda r 
+ \frac{3}{2r} \left( \frac{4}{3}-\frac{2m}{r} \right).
\label{e42}
\end{equation}
With the help of  Eqs.~(\ref{e40}), (\ref{e41}), and (\ref{e35}), 
we reach  
\begin{equation}
\frac{d}{dr}\left(Ee^{-\delta}\right)=
\left[ \Lambda r 
- \frac{3}{2r} \left( \frac{4}{3}-\frac{2m}{r} \right) \right]
e^{-\delta}\phi'^2.
\label{e43}
\end{equation}
Noting that $\Lambda < 0$, it is obvious  that the right hand side 
of Eq.~(\ref{e43}) is  always negative  for $r>r_b$, where $r_b$
represents the location of the black hole horizon.  Therefore the 
function  $Ee^{-\delta}$  should   be a   decreasing  function   for 
$r>r_b$. From Eq.~(\ref{e40}) we also note that 
$E(r_b)=-V[\phi(r_b)] < 0$.  Therefore $E  e^{-\delta} (r>r_b)$ is 
always more  negative than  $E  e^{-\delta} (r_b)$.  However the 
asymptotically anti-de Sitter solution satisfying the asymptotic 
behavior (\ref{e37}) requires that 
$\lim_{r \rightarrow \infty} Ee^{-\delta}  = 0$. Hence the only solution 
is $\phi'=0$ and $V(\phi)=0$ throughout the spacetime. In fact, the 
constant scalar can correspond to a local extremum of the potential
satisfying $\partial V(\phi)/\partial \phi =0$, because one can absorb 
the extremum of the potential to the cosmological constant so that the
cosmological constant is an effective one and the potential becomes an 
effective potential. Thus the above proof continues to hold. We will
discuss this point later.

(ii) $\Lambda > 0$.  This is another story due to the different 
asymptotic behavior.
On the one hand, the cosmological horizon may appear. On
the other hand, obviously, the right hand side of Eq.~(\ref{e43}) 
is positive semidefinite as $r$ is larger than a critical
value. Thus we have no way to rule out the possibility 
of having  scalar hairs in this method. If  defining 
\begin{equation}
E(r) = \frac{1}{2} \mu \phi'^2 - V( \phi ) - \frac{1}{4\pi} \Lambda,
\label{e44}
\end{equation}
we have
\begin{equation}
\frac{d}{dr}\left(Ee^{-\delta}\right)=
- \frac{3}{2r} \left( \frac{4}{3}-\frac{2m}{r} \right) 
e^{-\delta}\phi'^2.
\label{e45}
\end{equation}
The right hand side of Eq.~(\ref{e45}) now  is always negative 
for $r> r_b$. This indicates that the function $Ee^{-\delta}$ should
be a decreasing function for $r>r_b$. From (\ref{e44}) we  note 
that
$E(r_b) = -V[\phi(r_b)]-\Lambda / 4\pi < 0$ and
$E(r_c)=-V[\phi(r_c)]- \Lambda / 4\pi  <0 $. 
Therefore $Ee^{-\delta}(r>r_b)$ is  always negative semidefinite for
$r\ge r_b$.
Note that 
$\lim_{r \rightarrow \infty} E e^{-\delta} =- \Lambda  / 4\pi $
in contrast to the cases of 
the asymptotically  flat and asymptotically anti-de Sitter spacetimes
where $\lim_{r \rightarrow \infty} E e^{-\delta} = 0$.
Thus, for asymptotically de Sitter black holes,
we still have no reason to rule out the scalar hair 
and hence the Sudarsky's method does not work. 
In fact, both the Bekenstein's method proving his novel 
no scalar hair theorem~\cite{Bek} and 
the scaling techniques~\cite{Heu} do not work 
in this case, either. 
But by using Bekenstein's identity method~\cite{Bek1}
we can show the following theorem.

{\it Theorem 3}: In the   Einstein-minimally coupled   scalar field  
system with  a positive semidefinite, convex  scalar potential and a 
positive cosmological constant, the spherically symmetric, 
asymptotically de Sitter solution (with a cosmological horizon and
no black hole horizon) and the spherically symmetric, asymptotically 
de Sitter black hole solution 
(with a regular black hole horizon and a cosmological horizon) 
 are  the  de Sitter space and the Schwarzschild-de  Sitter black hole, 
 respectively.  And the scalar field is a  constant corresponding to the 
 point of the  minimum  of this potential.

{\it Proof}: Multiplying (\ref{e34}) by $(\phi -\phi _0)$ and 
integrating from $r_0$ to $r_c$, we have
\begin{eqnarray}
&& r^2e^{-\delta}\mu (r) \phi'(\phi -\phi_0) |^{r_c}_{r_0}
   \nonumber \\
&&~~~~~~~~=\int ^{r_c}_{r_0}dr \  r^2e^{-\delta}\left (\mu (r)\phi'^2
+(\phi -\phi _0)\frac{\partial V}{\partial \phi}\right ).
\label{e46}
\end{eqnarray}
Here $\phi _0$ is a constant 
which gives the minimum value of the potential $V$,
$r_c$ denotes the cosmological horizon and 
$r_0$ represents the origin  for the asymptotically de Sitter solution 
and black hole horizon $r_b$ for de Sitter black holes. 
The left hand side  of equation (\ref{e46}) always vanishes: 
For the black hole solution, 
the metric function $\mu =0$ both at the black hole 
and cosmological horizons; 
for the asymptotically de Sitter solution, $\mu=0$ at the cosmological 
horizon and $\phi'=0$ at the origin, which comes from the regularity
requirement.   If the scalar potential $V(\phi)$ is convex, 
both the two terms of integrand of  Eq.~(\ref{e46}) are 
then  positive semidefinite 
in the region between $r_0 \le r \le r_c$. Therefore the only 
solution is $\phi =\phi _0 $, and the metrics  are the Schwarzschild-de 
Sitter and the de Sitter space (as a special case of the 
Schwarzschild-de Sitter metric), respectively.

When the solution is an asymptotically anti-de Sitter one, if we set
that $r_0$ is the black hole horizon and $r_c$ is  asymptotic 
infinity, then the left hand side of Eq.~(\ref{e46}) still vanishes:
$\mu(r)=0$ at the horizon and $\phi'=0$ at infinity. And the integrand 
of Eq.~(\ref{e46}) is still positive semidefinite. Thus  the only
solution is the Schwarzschild-anti-de Sitter spacetime with a constant 
scalar field. In a word, 
the Bekenstein's identity method can exclude the scalar hair of
spherically symmetric (anti-)de
Sitter black holes and the de Sitter space if the scalar potential is 
convex. In fact, this method can also exclude 
the scalar hair for the Kerr-(anti-)de Sitter black holes.

\section{Conclusion and discussion}

In this work we have investigated the possibility of having 
hairs on the cosmological horizon and the effects of  
 a cosmological constant on the no short hair conjecture and
no scalar hair theorem of black holes. 
 From the viewpoint of having hairs on horizons, 
the cosmological horizon shares 
similar properties of the black hole horizon. 
For the theories admitting haired black hole solutions, 
the nontrivial matter configurations may reach  
and extend beyond the cosmological horizon.
An explicit example is that in the Bartnik-McKinnon soliton
surrounded by a cosmological horizon
the Yang-Mills field can extend beyond the cosmological horizon.
This has been found  numerically in Refs.~\cite{Torri,Volkov}.
For Q-stars and boson stars, the matter fields cannot reach 
the cosmological horizon.
The no short hair conjecture of black holes  remains valid, 
in spite of the different asymptotic behaviors (de Sitter or anti-de
Sitter) of black hole solutions in the sense that  the presence of the
cosmological constant does not change the expression of the 
critical point (\ref{e22}). But we would like to point out that,
 although the NQS's no 
short hair conjecture remains valid 
for asymptotically (anti-)de Sitter black holes,
we cannot say that the hair must extend 
beyond 3/2 the horizon radius of black holes,
since in our case the horizons are determined 
by the  equation, $1-2m(r)/r-\Lambda r^2 /3 =0$.
 The positive 
cosmological constant widens the hairosphere, while  
the hairosphere thins for a negative cosmological constant, compared to
the one  in the asymptotically flat black hole solutions
\cite{NQS}.

For a negative cosmological constant, we have shown:
For the Einstein-minimally coupled scalar field system 
with a positive semidefinite scalar potential,
if the spacetime is static and spherically symmetric, 
has regular black hole horizon, and is of the asymptotically 
anti-de Sitter behavior, the only solution  is the 
Schwarzschild-anti-de Sitter spacetime and the scalar field is a 
constant corresponding to a local extremum  of the potential.
For a positive cosmological constant, if the scalar potential is 
convex, both the cosmological horizon
and black hole horizon cannot support the scalar hair, that is, 
 the only solution is the de Sitter space or 
Schwarzschild-de Sitter spacetime. Therefore the no scalar hair theorem
hold not only for the asymptotically flat black holes, but also for the
asymptotically (anti-) de Sitter black holes.

 It is a  difficult task to prove no scalar hair theorem of black holes
which are not necessarily spherically symmetric.  By now one can prove 
the no scalar hair theorem for asymptotically flat black holes 
only when the scalar potential is convex.  When the scalar potential is 
non-negative, in order to prove the no scalar hair theorem
one must have more conditions: The spacetime is static, spherically 
symmetric.   When the spacetime is asymptotically anti-de Sitter or de 
Sitter, certainly, it will become more difficult to show the no scalar 
hair theorem.  In Sec. IV we have  proven the no scalar hair theorem 
of anti-de Sitter black holes in the case for a minimally coupled scalar 
field with a positive semidefinite scalar potential and
the metric being of form (\ref{e2}). 
For the asymptotically de Sitter solution  and de Sitter black hole,
however, it should be emphasized that when 
the scalar potential is convex, the result is always valid, in spite 
of whether or not  the cosmological constant is present and whether 
or not the spacetime is spherically symmetric.

Here it should also be stressed that 
a {\it general} scalar potential which is {\it bounded from below}
can be redefined to be positive semidefinite
by absorbing the negative value into the cosmological constant.
Then our cosmological constant is an effective one 
and it determines
the asymptotic behavior of the spacetime. So in the sense of 
no scalar hair theorem we have shown:
In the Einstein-minimally coupled scalar field system with an arbitrary 
scalar potential, the only spherically symmetric black hole 
solution is the Schwarzschild-anti-de Sitter spacetime if the effective 
cosmological constant is negative, or the Schwarzschild-de Sitter
spacetime if the effective cosmological constant is positive and
the effective potential is convex.  
The scalar field is a trivial constant corresponding to a local
extremum of the potential 
[$ \partial V(\phi)/\partial \phi =0$]. 
Since the hair is a characteristic of black holes, it may be 
closely related to the thermodynamics of black holes~\cite{Bek2}. 
Therefore it is significant to study under what conditions black  
holes have hairs, and under what conditions 
black holes have no hairs for asymptotically flat black holes, 
asymptotically (anti-)de Sitter black holes, and
even for black holes with unusually asymptotic behavior and
non-spherically topological  black holes.  
In addition, it also should be of interest 
to verify numerically that the cosmological horizon can
support the nontrivial matter hairs 
in those theories admitting haired black hole solutions.

\section*{Acknowledgments}

This work was supported by 
the Center for Theoretical Physics (S.N.U.) 
and by the Non-Directed Research Fund, Korea Research Foundation, 1996.
R.G.C. would like to thank M. Heusler for helpful correspondence 
and for  sending new book in Ref.~\cite{Heu} to him, 
Profs.  S. P. Kim, C. K. Lee, K. S. Soh and H. S. Song 
for a great deal of kind help, and Dr. J. H. Cho for  many 
stimulating discussions. We also thank the referee for informing us
of many related references and useful suggestions which helped 
us to improve this paper.

\end{document}